\documentclass[twocolumn,aps,prx,sermicityowpacs,amsmath,superscriptaddress,longbibliography,notitlepage]{revtex4-2}
\usepackage{times}
\usepackage{amssymb}
\usepackage{mathrsfs}
\usepackage{graphicx}
\usepackage{float}
\usepackage[caption=false]{subfig}
\usepackage[colorlinks,bookmarks=true,citecolor=blue,linkcolor=blue,urlcolor=blue]{hyperref}
\usepackage{verbatim}
\usepackage{epstopdf}
\usepackage{verbatim}
\usepackage{xcolor}
\usepackage{bm}
\usepackage{soul}
\usepackage{ulem}

\newcommand{\clr}{\color{red!75!black}}

\begin{document}

\title{Topologically Protected Exceptional Points and Reentrant $\mathcal{PT}$ Phase in an Exact Ternary Model}

\author{Chulwon Lee}
\thanks{These authors contributed equally}
\affiliation{Department of Physics, University of Michigan Ann Arbor, Ann Arbor, Michigan, 48109, United States}
\author{Kai Zhang}
\thanks{These authors contributed equally}
\affiliation{Department of Physics, University of Michigan Ann Arbor, Ann Arbor, Michigan, 48109, United States}
\author{Jinyan Miao}
\affiliation{Department of Physics, University of Michigan Ann Arbor, Ann Arbor, Michigan, 48109, United States}
\author{Kai Sun}\email{sunkai@umich.edu}
\affiliation{Department of Physics, University of Michigan Ann Arbor, Ann Arbor, Michigan, 48109, United States}
\author{Hui Deng}\email{dengh@umich.edu}
\affiliation{Department of Physics, University of Michigan Ann Arbor, Ann Arbor, Michigan, 48109, United States}

\begin{abstract}
In open, driven systems where parity-time symmetry is preserved, phenomena that defy conventional wisdom emerge near exceptional points, promising advances in photonics. While most studies focus on two-level systems of a conventional exceptional point, unconventional exceptional points as well as reentrant phases have been discovered in separate studies of higher-dimensional phase spaces. 
In this Letter, we present a minimal, analytical model that encompasses several key phenomena in higher-dimensional phase spaces, including reentrant $\mathcal{PT}$ phases, higher-order exceptional points, and anisotropic exceptional points. 
Using the exact analytical solution, we identify a new topological index as the unifying origin of these different phenomena. 
The simplicity of the model may furthermore facilitate experimental implementations for enhanced sensing and efficient polariton devices. 
\end{abstract}

\maketitle

\noindent\textit{\clr Introduction.}
Most quantum physics models consider closed, Hermitian systems as an approximation to real physical systems. 
Following the seminal work by Bender and Boettcher~\cite{BenderPT1998}, it has been found that non-Hermitian systems with parity-time ($\mathcal{PT}$) symmetry feature many experimentally relevant novel phenomena that are absent in Hermitian models~\cite{Bender1999,Bender_2007,Ashida2020}. 
These phenomena are often associated with the $\mathcal{PT}$ phase transition at exceptional points (EPs), singularities in phase space where the modes coalesce.
In two-level photonic systems, EPs have been demonstrated with optical gain or loss as control parameter in a 1D phase space~\cite{Regensburger2012,Chen2017Nature,Miri2019,YangLan2019}, and have enabled improved sensing, lasing, and optical signal transport~\cite{Liertzer2012PRL,MohammadSci2019,HanScience2019}.
Further studies led to a deeper understanding of the nature of EPs through their topological classifications~\cite{shen2018topological,kawabata2019classification,delplace2021symmetry,mandal2021symmetry,tang2023realization,zhang2023symmetry,rivero2022imaginary}. 
More recently, it has been shown that higher-dimensional phase space, most readily realized with multilevel systems, can host much richer phenomena with further improved performance in various applications, such as anisotropic EP (aEP), with anisotropic parameter-dependent dispersion~\cite{shen2018topological,ding2018experimental,xiao2019anisotropic,jin2020hybrid,Tang2020}, and higher-order EP (hEP), where multiple modes coalesce ~\cite{DingKun2016PRX,Hodaei2017Nature,ShenHT2018PRL,DingKun2018PRL,xiong2022higher,tang2023realization}. A particularly counterintuitive feature is the reentrant $\mathcal{PT}$ exact phase~\cite{Nicolas2015}. While typically stronger non-Hermitian effects, such as stronger damping, lead to the $\mathcal{PT}$ broken phase, some multi-level systems can reenter the $\mathcal{PT}$-exact phase as the non-Hermitian effects become even stronger.

When do phase reentrance and the new types of EPs take place? What are their topological properties?
Are there fundamental connections between these diverse, separately identified phenomena in multi-level non-Hermitian systems?

\begin{figure}[b]
    \begin{center}		 
    \includegraphics[width=0.95\linewidth]{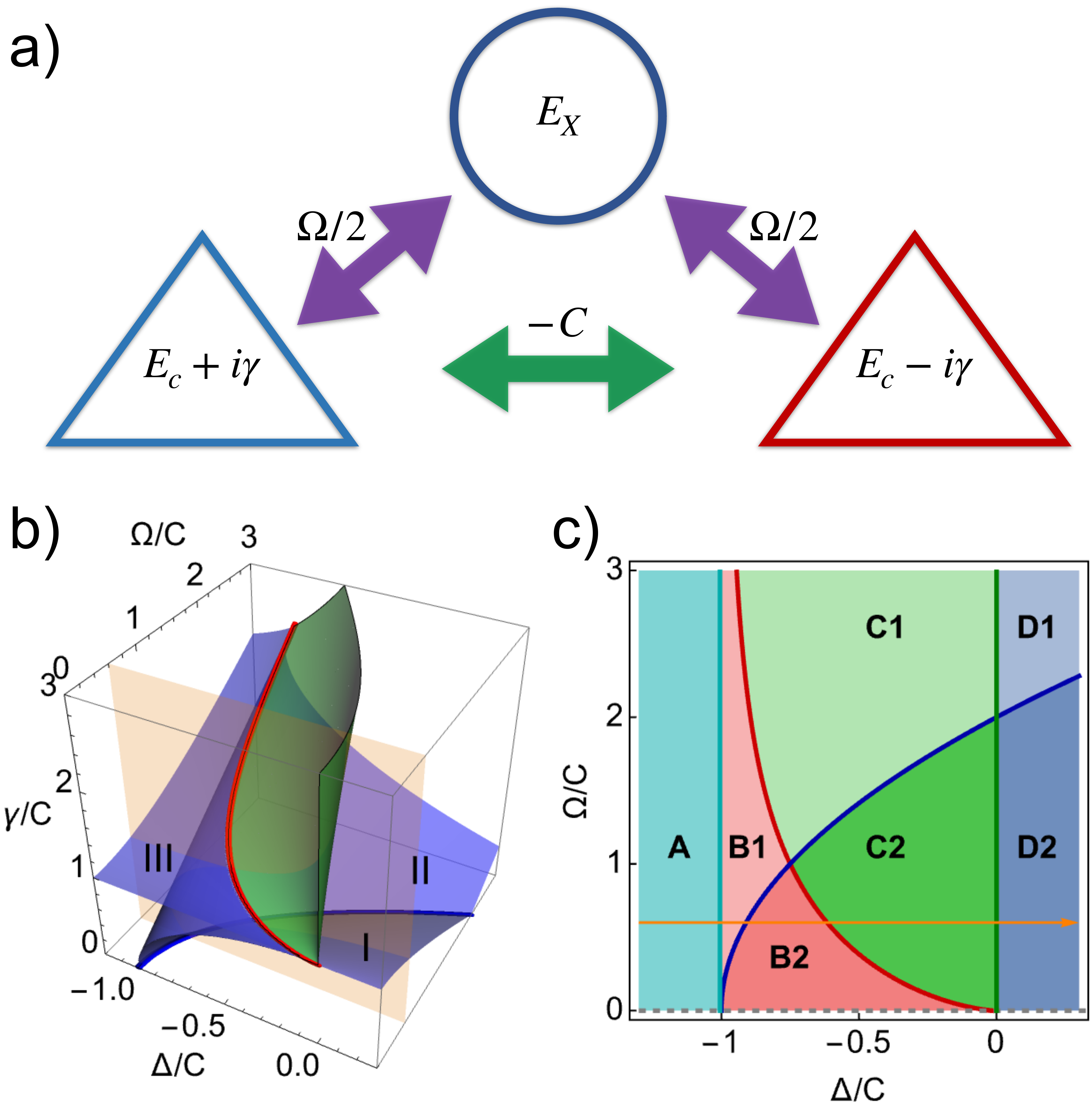}
    \par\end{center}
    \protect\caption{\label{fig:1} 
    Schematic description of a non-Hermitian exciton-polariton system and its full phase diagram in the parametric space. a) Schematic to minimal non-Hermitian exciton-polariton model; (b) The full phase diagram of the 3$\times$3 non-Hermitian polariton system; (c) the projection of the 3D phase diagram onto the $\tilde{\Delta}-\tilde{\Omega}$ plane.} 
\end{figure}

In this work, we present a minimal $\mathcal{PT}$ symmetric non-Hermitian model that hosts these key features in higher-dimensional systems -- reentrant $\mathcal{PT}$ exact phase, aEP, and hEP; and we furthermore analyze the parity and topological connections of these phenomena.
We obtain analytically the full phase diagram of our minimal model and show that the $\mathcal{PT}$-reentrant phase results from a polariton-driven, curved $\mathcal{PT}$ phase boundary, terminating at aEP and hEP. 
By defining a generalized parity indicator, we show that the reentrant phase can be divided into two types, depending on whether there is a parity inversion upon reentrance.
Using the analytically solved phase diagram, we identify a complex vector field where aEP and hEP are two topologically protected singularities with opposite topological charges; 
the topological charges vanish only when aEP and hEP coalesce, corresponding to the disappearance of the reentrant $\mathcal{PT}$ phase. 

Our model provides a unified framework for understanding the rich features and inherent topology of higher-dimensional $\mathcal{PT}$ symmetric non-Hermitian systems. 
The multiple degrees of freedom of such systems also allow tuning and control for applications exploiting the new types of EPs. 

\begin{figure*}[t]
    \begin{center}
    \includegraphics[width=0.9\linewidth]{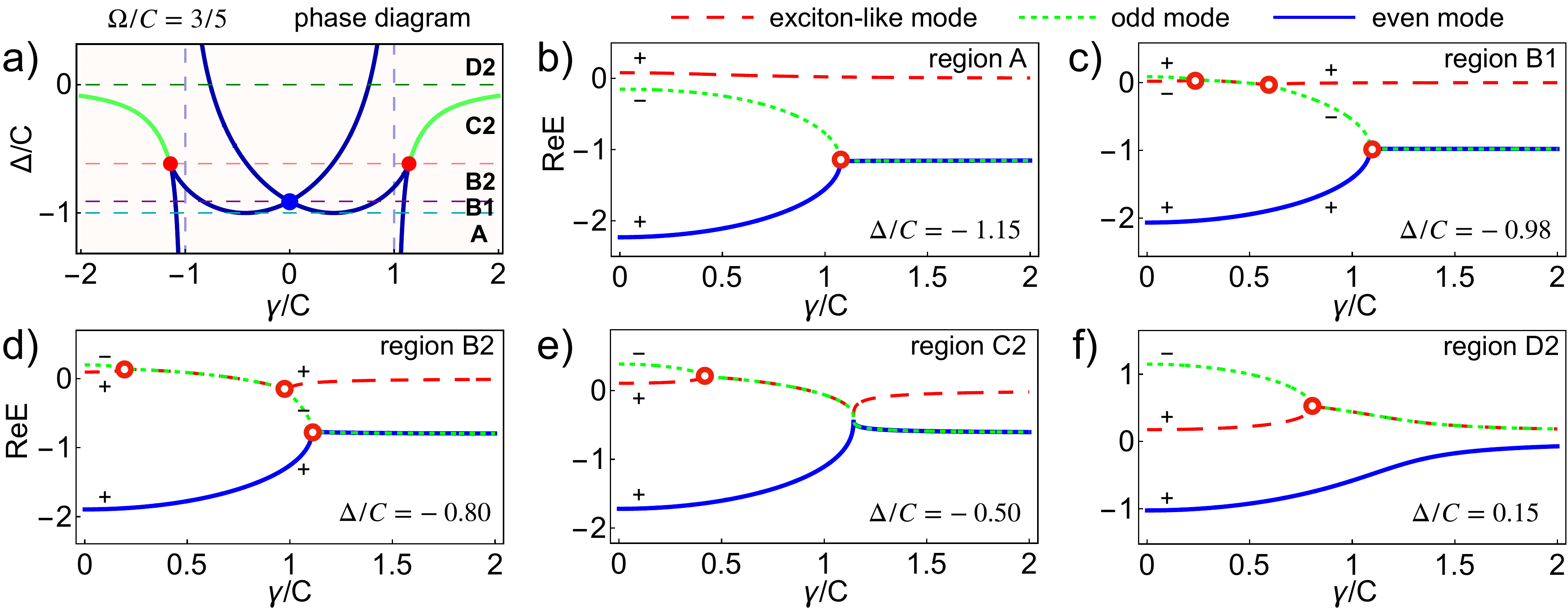}
    \par\end{center}
    \protect\caption{\label{fig:2} 
        (a) The 2D phase diagram for the $\tilde{\Omega}=3/5$ slice, which corresponds to the orange plane in Fig.~\ref{fig:1}(b) or the path indicated by the orange arrow in Fig.~\ref{fig:1}(c). The blue dashed line indicates the $\mathcal{PT}$ phase boundary when $\tilde{\Omega}=0$. (b-f) The real parts of the three eigenvalues evolve with gain and loss $\tilde{\gamma}$ under different cavity detuning $\tilde{\Delta}$. Here, the red circles indicate the exceptional points, and the Rabi coupling strength $\tilde{\Omega}$ is always chosen to be $3/5$.} 
\end{figure*}

\textit{\clr Minimal $\mathcal{PT}$-symmetric model.}
To obtain phase reentrance in a linear, non-Hermitian $\mathcal{PT}$-symmetric system, a minimum of three levels are required (See~Appendix~\ref{AppendixA}). We construct here a minimal three-level system as illustrated in Fig.~\ref{fig:1}(a).  It consists of two coupled $\mathcal{PT}$ symmetric modes, which we call cavities, simultaneously coupled to a third lossless mode, which we call an exciton.
The two cavities have degenerate resonance energy $E_c=\hbar\omega$ and balanced gain and loss $\gamma$; they are coupled with each other with the coupling strength $C$. 
The exciton mode has a resonance energy $E_X$, detuned from the cavities by $\Delta=E_c-E_X$. It is lossless and couples to both cavities with the same coupling strength $\Omega/2$. 
The Hamiltonian of the system can be expressed as
\begin{equation}\label{MT_EffHam1}
H = 
\begin{pmatrix}
E_c-i \gamma & -C & \Omega/2 \\ 
-C & E_c+i \gamma & \Omega/2 \\
\Omega/2 & \Omega/2 &E_X
\end{pmatrix},
\end{equation}
where the negative sign of coupling strength is chosen such that the lower (higher) frequency coupled mode possesses even (odd) parity for $C>0$. 

The Hamiltonian is a minimal ternary model that is invariant under $\mathcal{PT}$ operation where coupled-cavity exciton-polariton is described. 
The parity operator $\mathcal{P}$ of the system has the representation
\begin{equation}\label{MT_ParityOpt}
\mathcal{P} = 
\begin{pmatrix}
    0  & 1 & 0 \\ 
    1  & 0 & 0 \\ 
    0 & 0 & 1 \\ 
\end{pmatrix}
\end{equation}
and the time-reversal operator $\mathcal{T}$ refers to the conjugate complex on this physical basis. 
To simplify the notation, we rewrite the Hamiltonian as $\tilde{H} = \frac{1}{C}(H - E_X \mathbb{I}_3)$, where $\mathbb{I}_3$ represents the $3\times 3$ identity matrix, and we redefine parameters normalized by the cavity coupling constant $C$ as $\tilde{\Delta}=\Delta/C$, $\tilde{\gamma}=\gamma/C$ and $\tilde{\Omega}=\Omega/C$. The effective Hamiltonian can be written as
\begin{equation}\label{MT_EffHam3}
\tilde{H}= 
\begin{pmatrix}
    \tilde{\Delta}-i \tilde{\gamma} & -1 &\tilde{\Omega}/2 \\
    -1 & \tilde{\Delta}+i \tilde{\gamma} & \tilde{\Omega}/2\\
    \tilde{\Omega}/2 & \tilde{\Omega}/2 & 0
\end{pmatrix}.
\end{equation}
The eigenvalue $E$ and the corresponding eigenvector $\psi$ satisfy the eigen equation $\tilde{H}\psi=E\psi$. 

\textit{\clr Solving for the phase diagram of the minimal model.}~The phase diagram of a $\mathcal{PT}$ symmetric Hamiltonian consists of regions separated by phase boundaries where degeneracy of the energy levels occurs. For our ternary system, its phase boundaries can be solved analytically using the characteristic equations of the Hamiltonian~\cite{woody2016polynomial,Discriminant_Notes,SupMat} as we describe below. The resulting full phase diagram is shown in Fig.~\ref{fig:1}(b). 

The main phase boundary is where two or more eigenvalues coalesce, with the same real and imaginary parts. These points correspond to exceptional points (surfaces) in a 1D (3D) parameter space. They form a $\mathcal{PT}$ phase boundary, separating the $\mathcal{PT}$ exact and $\mathcal{PT}$ broken phases. We solve for EP by analyzing the discriminant of the characteristic equation $f$ of the matrix in Eq.~\ref{MT_EffHam3}: 
$f(E, \tilde{\Delta}, \tilde{\Omega},\tilde{\gamma}) = \mathrm{det} [E \, \mathbb{I}_3-\tilde{H}] = 0$, or,
\begin{equation}\label{MT_CharacEqu}
f=E^3-2 \tilde{\Delta} E^2+
   (\tilde{\Delta}^2+\tilde{\gamma}^2-\tilde{\Omega}^2/2-1)E \\ 
    +(\tilde{\Delta}+1)\tilde{\Omega}^2/2=0.
\end{equation}
The characteristic equation is a real polynomial in energy $E$, due to the $\mathcal{PT}$ symmetry of the Hamiltonian, with coefficients governed by system parameters.
Therefore, the system resides in the $\mathcal{PT}$ exact phase when the discriminant is positive: $\mathrm{Disc}_E(f)>0$, yielding three real eigenvalues. 
Conversely, when $\mathrm{Disc}_E(f)<0$, the system enters the $\mathcal{PT}$ broken phase, featuring complex eigenvalues $E$, with two forming a complex conjugate pair and the third remaining real.
The boundary between the $\mathcal{PT}$ exact and broken phases is therefore determined by
\begin{equation}\label{eq:phaseboundary}
    \mathrm{Disc}_E(f)=0,
\end{equation}
where a pair of the eigenvalues coalesce, forming EP surfaces, shown as the blue surfaces in Fig.~\ref{fig:1}(b). 

In the $\mathcal{PT}$ broken phase, there exists another type of degeneracy, where all eigenvalues have the same real part but different imaginary parts. We call these points degenerate points (DG). They can be determined with the characteristic equation Eq.(\ref{MT_CharacEqu}) using the resultant method~\cite{SupMat} and corresponds to the green surface in Fig.~\ref{fig:1}(b). 

Note that the characteristic equation is invariant if we flip the signs of $\tilde{\Omega}$ and $\tilde{\gamma}$. 
Therefore, the 3D phase diagram is symmetric with respect to the $\tilde{\Omega}=0$ and $\tilde{\gamma}=0$ planes. 
It is sufficient to consider the regions $\tilde{\Omega}\geq 0$ and $\tilde{\gamma}\geq 0$ of the phase diagram. 

As shown in the phase diagram in Fig.~\ref{fig:1}b, the system features three smooth EP surfaces (blue surfaces I, II and III), which give rise to reentrant $\mathcal{PT}$ phase. At the same time, the three EP surfaces intersect at two special creases, labeled as level-crossing line (purple curve) and hEP line (red curve). The middle EP surface II has a concave shape, leading to another special EP line, labeled as aEP line. In addition, the green DG surface is bound by the asymptotic line of $|\tilde{\gamma}|\rightarrow\infty$, which requires $\lim_{\tilde{\gamma}\rightarrow {+\infty}}\partial_{\tilde{\Delta}}\tilde{\gamma} = -\partial_{\tilde{\Delta}}f/\partial_{\tilde{\gamma}}f = (2E-2\tilde{\Delta}- \tilde{\Omega}^2/2E)/2\tilde{\gamma}=0$, or, $\tilde{\Delta}=0$ for non-zero $\tilde{\Omega}$. 
Below we first analyze the $\mathcal{PT}$ phase transitions, then the features of the special EP lines. 

\textit{\clr Analysis of the phase diagram.}~The multiple EP surfaces form a three-fold covering over a region in the $\tilde{\Delta}-\tilde{\Omega}$ plane, which leads to reentrant $\mathcal{PT}$ phase transitions with increasing $\tilde{\gamma}$. 
Such regions can be identified more clearly by projecting the 3D phase diagram onto the $\tilde{\Delta}-\tilde{\Omega}$ plane. As shown in Fig.~\ref{fig:1}(c), this region with $\mathcal{PT}$ phase reentrance is bounded by the projected hEP and aEP lines, labeled as region $B$, while other regions feature a single $\mathcal{PT}$ phase transition. The level-crossing line at $\tilde{\gamma}=0$ and the asymptotic line at $\tilde{\Delta}=0$ further divide the projected phase diagram to a total of seven regions, with different numbers of exceptional points, real eigenvalue degeneracy, or mode parities. 

To illustrate how the system evolves with $\tilde{\gamma}$ in these different regions, we examine the cross section of the 3D phase diagram at a given $\tilde{\Omega}$. An example is shown in Fig.~\ref{fig:2}(a) for $\tilde{\Omega}=3/5$, marked by the orange plane in Fig.~\ref{fig:1}(b). 
The blue and green lines are the intersections of the orange plane with the blue and green surfaces, respectively, in Fig.~\ref{fig:1}(b).
The blue EP line separates the $\mathcal{PT}$ exact and broken phase. 
The evolution of the system with $\tilde{\Delta}$ follows the orange arrow in Fig.~\ref{fig:1}(c), crossing $A,B1,B2,C2$ and $D2$ regions, separated by the vertical dashed lines in Fig.~\ref{fig:2}(a). 
The five regions possess different numbers of phase boundaries along the path of increasing $\tilde{\gamma}$. 
In regions $A$, $C1$ and $D2$, for a given $\tilde{\Delta}$, the system passes the EP boundary once with increasing $\tilde{\gamma}$, while it crosses EP boundaries three times in the regions $B1$ and $B2$. 
That is, the system undergoes a transition from $\mathcal{PT}$ exact to $\mathcal{PT}$ broken phase, then reenters the $\mathcal{PT}$ exact phase, and eventually transitions again to the $\mathcal{PT}$ broken phase at large gain/loss $\tilde{\gamma}$.

Note that the reentrant $\mathcal{PT}$ exact phase is driven by the polaritonic coupling $\tilde{\Omega}$ in the ternary system. When $\tilde{\Omega}$ vanishes, the system is reduced to a coupled-cavity system of two levels. 
The system undergoes the $\mathcal{PT}$ phase transition only once when $\gamma=\pm 1$, shown as the blue vertical dashed lines in Fig.~\ref{fig:2}(a) for reference, and no phase reentry occurs.  

Accompanying the different evolution of the $\mathcal{PT}$ phases, the eigenstates of the system also evolve differently. We show in Fig.~\ref{fig:2} examples of the real part of the eigenvalue $\mathrm{Re}(E)$ of each eigenstate versus $\tilde{\gamma}$ in each region. The EPs are indicated by the red circles. 
In region $A$ (Fig.~\ref{fig:2}(b)), where the detuning $\Delta/C<-1$, there is a single EP and the two cavities behave like a two-level $\mathcal{PT}$ symmetric non-Hermitian system, while the exciton is largely uncoupled (Fig. S1(b)).
In regions $B1$ and $B2$ (Fig.~\ref{fig:2}(c)-(d)), at a smaller negative detuning, $\mathcal{PT}$ phase reentrance takes place. 
As shown by the eigenvalue evolution, the upper EP surface III at relatively large $\gamma$ corresponds to coalescing of the two cavity modes,(Fig.~\ref{fig:2}(b)) and it exists only when the cavity energy is lower than the exciton energy ($\Delta/C\leq 0$). In contrast, the middle and lower EP surfaces, II and I, both correspond to coalescing of the exciton and one cavity, forming polaritonic modes (Fig. S1(c-d)).  
In region $C2$ (Fig.~\ref{fig:2}(e)), at even smaller negative detuning, there are no longer reentrant $\mathcal{PT}$ phases. 
Lastly, in region $D2$ with positive detuning ($\Delta/C>0$) (Fig.~\ref{fig:2}(f)), there is also only one EP. 
The eigenstates evolve from three non-degenerate coupled modes to a pair of degenerate cavity-like modes and an exciton-like mode.

\begin{figure*}[t]
\begin{center}
\includegraphics[width=.85\linewidth]{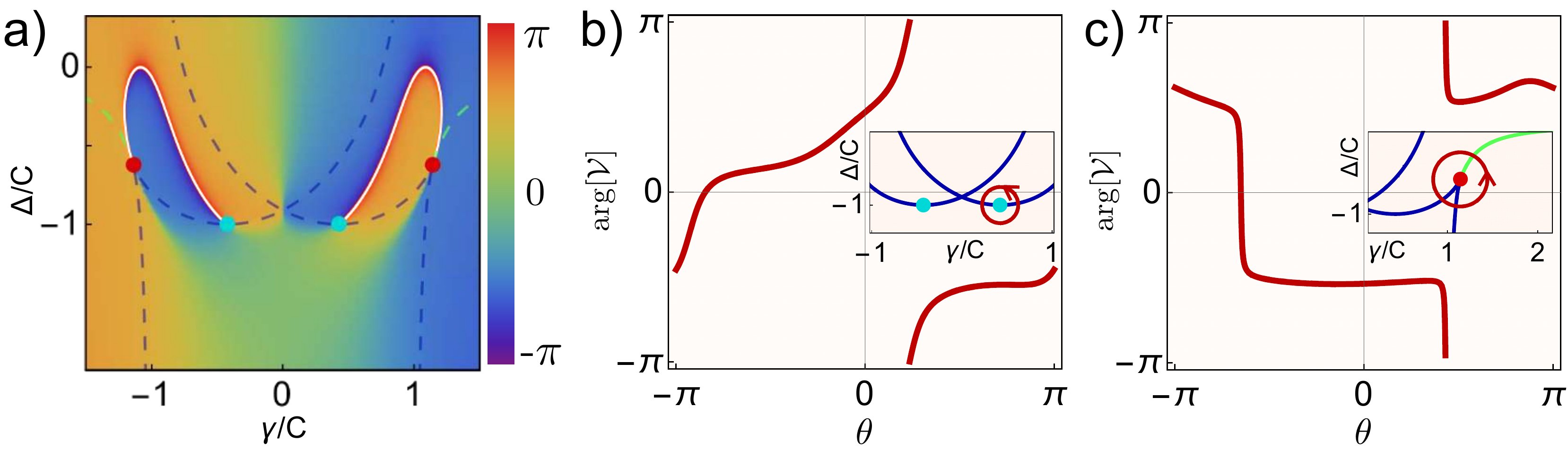}
\par\end{center}
\protect\caption{\label{fig:3} 
    Topological classification of anisotropic and higher-order exceptional points. 
    (a) Vector field $\mathcal{V} = \mathrm{Disc}_E(f) + i \, \partial_{\tilde{\gamma}}\mathrm{Disc}_E(f)$ mapped on the phase diagram. (dashed line) Vortices corresponding to aEP and hEP are indicated by blue and red dots, respectively. Note that the winding numbers for aEP and hEP are given opposite.
    (b) and (c) are argument function of the vector field $\mathcal{V}$ along the contour encircling aEP and hEP, respectively. (insets)
    }
\end{figure*}

\textit{{\clr Generalized parity indicator.}} Parity is no longer a good quantum number for non-Hermitian systems. 
Instead, we introduce a generalized parity indicator for the $\mathcal{PT}$ exact phases to analyze the parity symmetry.
We define the generalized parity indicator as 
\begin{equation}\label{MT_ParityInd}
	\epsilon_n = \frac{\langle\psi_n^L|\mathcal{P}|\psi_n^R\rangle}{|\langle\psi_n^L|\mathcal{P}|\psi_n^R\rangle|},
\end{equation}
where the superscript $R,L$  represent the right and left eigenvectors of the Hamiltonian, respectively~\cite{SupMat}.
For eigenstates in the $\mathcal{PT}$ exact phase, $\epsilon_n = \pm 1$ (See~Appendix~\ref{AppendixB}), as labeled by $\pm$ signs in Fig.~\ref{fig:2}.

At small $\tilde{\gamma}$, the parity of the two highest eigenstates are inverted between region 1 (Fig.~\ref{fig:2}(b)(c)) and region 2 (Fig.~\ref{fig:2}(d)(e)(f)). 
The system can evolve adiabatically between different $\mathcal{PT}$ phases when the parity indicator is invariant for each mode, such as within $B1$ (Fig.~\ref{fig:2}(b)). 
However, when the parity indicator of the modes possesses opposite signs in different phase regions, adiabatic evolution between them is prohibited. 
This is the case for the first and second $\mathcal{PT}$ exact phases in region $B2$ (Fig.~\ref{fig:2}(c)), at small and large $\tilde{\gamma}$, respectively; so the two phases are always separated by a $\mathcal{PT}$ broken phase. Similarly, adiabatic evolution is also prohibited between the $\mathcal{PT}$ exact phase at small $\tilde{\gamma}$ in regions $B1$ and $B2$  (Fig.~\ref{fig:2}(b) and (c)). 

The change in the generalized parity indicator between the B1 and B2 regions originates from the level-crossing (LC) line, where the EP blue surface intersects with the $\tilde{\gamma}=0$ plane (Fig.\ref{fig:1}(b)). 
When the system crosses the LC line, the parity of two eigenstates will be exchanged. 
The LC also gives rise to an interesting feature, a near-zero-gain $\mathcal{PT}$ broken phase~\cite{ZhenBo2015}.
At LC, any finite $\tilde{\gamma}$ will drive the system immediately to the $\mathcal{PT}$ broken phase at LC, (Fig.~\ref{fig:2}(a)) which may be utilized for realizing ultra-low threshold photonic devices. The derivation of LC is discussed in the Supplementary Material~\cite{SupMat}.

\textit{\clr Anisotropic and higher-order EPs.}~
As illustrated in Fig.~\ref{fig:1}(c), the reentrant $\mathcal{PT}$ phases (B1 and B2 regions) are bounded by the cyan line and the red curve, corresponding to the aEP and hEP lines in the 3D phase diagram, respectively. 
The hEP line is composed of higher-order EPs, where all three eigenstates coalesce and the eigenspace collapses from three to one dimension. 
Such higher-order EP features enhanced sensitivity of $\delta E \approx \tilde{\gamma}^{1/3}$~\cite{Hodaei2017} as shown in~\cite{SupMat}. 
The aEP line is defined as the extremum of the EP surface II with respect to $\tilde{\gamma}$. 
The system exhibits different sensitivities as it passes through the aEP with increasing $\tilde{\gamma}$ or $\tilde{\Omega}$, respectively, that is, $\delta E\propto \delta \tilde{\gamma}$ and $\delta E\propto (\delta \tilde{\Delta})^{1/2}$. 

While regular EP corresponds to $\mathrm{Disc}_E(f)=0$ (Eq.~\ref{eq:phaseboundary}), aEP and hEP require an additional condition: 
\begin{equation}\label{eq:firstdbound}
    \partial_{\tilde{\gamma}} \mathrm{Disc}_E(f) = 0.
\end{equation}
This is because aEP is the extremum (the cyan dots in Fig.~\ref{fig:3}(a)), and hEP is a saddle point on the $\mathcal{PT}$ phase boundary (See~\cite{SupMat} and Appendix~\ref{AppendixC}). 

The aEP can also be viewed as coalescing two EPs in the reentrant $B1$ region into one by approaching $\tilde{\Delta}=-1$, which is analogous to a Dirac EP~\cite{jin2020hybrid} in a band system. 
As aEP is flat with respect to $\tilde{\gamma}$ and purely real, polariton modes at the proximity of this point can be long-lasting and are barely perturbed by $\tilde{\gamma}$. 
Furthermore, the parity of the two highest modes switches when they cross at aEP, which may be utilized as a polaritonic parity switch~\cite{SupMat}.

\textit{\clr Topological classification.}~
The mathematical constraints for aEP and hEP are useful for understanding the topological properties of aEP, hEP, and $\mathcal{PT}$ phase reentrance.
The left sides of the constraints in Eq.~\ref{eq:phaseboundary} and Eq.~\ref{eq:firstdbound} are both real functions of the system parameters, due to the $\mathcal{PT}$ symmetry of the Hamiltonian. Therefore we can use them to define a complex vector field:
\begin{equation}
    \mathcal{V} = \mathrm{Disc}_E[f] + i \, \partial_{\tilde{\gamma}}\mathrm{Disc}_E[f],
\end{equation}
and the winding number along a closed curve $C$ in this complex vector field:
\begin{equation}
    w = \frac{1}{2\pi}\oint_{C} ds \, \partial_s \arg \mathcal{V}(\tilde{\Delta},\tilde{\gamma}). 
\end{equation}

The vector field $\mathcal{V}$ is singular at aEP and hEP, but nonzero at regular EP (Fig.~\ref{fig:3}(a)).
Therefore, for $C$ that encircles aEP (inset of Fig.~\ref{fig:3}(b)) or hEP (inset of Fig.~\ref{fig:3}(c)) counterclockwise, the winding number measures how many times the vector field goes around the origin, giving a $\mathbb{Z}$-type topological classification of aEP and hEP (Fig.~\ref{fig:3}(b and c)). 
Remarkably, the winding numbers for aEP and hEP have opposite signs (Fig.~\ref{fig:3}(a)), showing that the net vortices can vanish only when aEP and hEP coalesce and the reentrant $\mathcal{PT}$ phase disappears. This condition can be satisfied only by $\tilde{\Omega} \rightarrow \infty$, and thus the reentrant $\mathcal{PT}$ phase is strongly protected by the topology of aEP and hEP.

\textit{\clr Summary.}~In conclusion, our ternary non-Hermitian system not only forms a minimal model with $\mathcal{PT}$ reentrant phase and corresponding aEPs and hEPs, but also allows full analytical solution of the phase diagram. The phase diagram, together with a generalized parity indicator and a vector field derived using analytical constraints of EPs, we reveals the parity symmetry origin of reentrant $\mathcal{PT}$ symmetry and its underlying connection with topological features unique to multimode systems -- near-zero-gain $\mathcal{PT}$ breaking and the anisotropic and higher-order EP lines that are Z-type topological singularities.

These unique features of multi-level $\mathcal{PT}$-symmetric non-Hermitian systems can be utilized for photonic technologies, such as enhanced sensitivity in sensing near hEP~\cite{de2022non,demange2011signatures,lin2016enhanced,jing2017high,Hodaei2017,xiong2022higher}, robust polariton modes near aEP that are persistent against changes in gain and loss, and near-zero-gain $\mathcal{PT}$ phase transitions near LC for energy-efficient polaritonic devices.

\begin{acknowledgements}
This work was supported in part by the Office of Naval Research Award N00014-21-1-2770 (CL, JM, KS and HD), the Gordon and Betty Moore Foundation Award GBMF10694 (CL, JM, KS and HD), the Army Research Office Award W911NF-17-1-0312 (CL, JM and HD), the National Science Foundation Award DMR 2132470 (KS and HD), and the Office of Naval Research Award MURI N00014-20-1-2479 (KZ and KS).
\end{acknowledgements}

\appendix
\setcounter{equation}{0}
\setcounter{secnumdepth}{2}

\section{Minimal linear non-Hermitian model with reentrant $\mathcal{PT}$ transition.}\label{AppendixA}
Here, we prove that our model serves as a minimal model of a linear, $\mathcal{PT}$-symmetric Hamiltonian that hosts the reentrant $\mathcal{PT}$ exact phase. 

We first consider a binary system. A generic linear non-Hermitian Hamiltonian can always be expressed in terms of Pauli matrices as:
$$H_{\mathrm{bi}} = a \sigma_0 + b \sigma_x + c \sigma_y + d \sigma_z$$
where $a,b,c,d$ are generally complex numbers. 
Imposing $\mathcal{PT}$ symmetry on this Hamiltonian leads to the condition:
\begin{equation}
    \sigma_x H_{\mathrm{bi}}^{\ast} \sigma_x = H_{\mathrm{bi}},
\end{equation}
which in turn requires $a, b, c$ to be real numbers, and $d$ to be an imaginary number. 
Now, we rewrite our binary model Hamiltonian that preserves $\mathcal{PT}$ symmetry as:
\begin{equation}
    H_{\mathrm{bi}} = a_r \sigma_0 + b_r \sigma_x + c_r \sigma_y + i d_r \sigma_z
\end{equation}
with $a_r,b_r,c_r,d_r$ real numbers. 
The eigenvalues of this system are $E_{\pm} = \pm \sqrt{a_r^2 +b_r^2 +c_r^2 -d_r^2} = \pm \sqrt{\Delta^2 -d_r^2}$, with $\Delta^2 = a_r^2 +b_r^2 +c_r^2$. 
When $|d_r|< |\Delta|$, the binary system lies in the $\mathcal{PT}$-exact phase. 
When the non-Hermitian parameter $d_r$ exceeds $\Delta$, the system enters $\mathcal{PT}$ broken phase. 
Further increasing the non-Hermitian parameter will not allow the system to reenter $\mathcal{PT}$ exact phase. 
We thus conclude that a linear, binary system preserving $\mathcal{PT}$ symmetry cannot exhibit the reentrance of the $\mathcal{PT}$ exact phase. Therefore, at least a ternary system is necessary.

We have shown in the main text that our teneray system preserves $\mathcal{PT}$ symmetry and hosts $\mathcal{PT}$ exact phase reentrance. Therefore, it forms the minimal model for reentrant $\mathcal{PT}$ exact phase.

\section{Analytic verification for 
generalized parity indicator in $\mathcal{PT}$ exact phase.}\label{AppendixB}
According to the phase diagram illustrated in Fig.~\ref{fig:1}(c), the region $B$ (red region) is further divided into two regions $B1$ and $B2$. 
The eigenvalue evolutions with the non-Hermitian parameter $\gamma$ in regions $B1$ and $B2$ are given in Fig.~\ref{fig:2}(c) and (d), respectively, which appear similar. 
To understand their inherent differences, it is necessary to generalize the mode parity indicator from Hermitian to non-Hermitian systems. 
In Hermitian systems with parity symmetry, the parity indicator of the $n$-th mode is well-defined as the eigenvalue of the parity operator $\mathcal{P}$ on this eigenmode, expressed as $\langle \psi_n| \mathcal{P} |\psi_n \rangle$. 
In non-Hermitian systems, the parity symmetry is broken due to $\tilde{\gamma}\neq 0$, while $\mathcal{PT}$ symmetry is retained. 
Therefore, we propose a generalized parity indicator that is also valid for the non-Hermitian $\mathcal{PT}$ exact phase, especially for the reentrant $\mathcal{PT}$ exact phase after a $\mathcal{PT}$ phase transition. 
The generalized parity indicator is defined as Eq.(\ref{MT_ParityInd}) in the main text. 
The biorthogonal basis satisfies 
\begin{equation}\label{AEQ_BiBasis}
   \tilde{H}|\psi^R_n\rangle=E_n|\psi^R_n\rangle; \quad 
   \tilde{H}^{\dagger} |\psi_n^L\rangle= E_n^{\ast} |\psi^L_n\rangle
\end{equation}
The generalized parity indicator has the value of either $\pm 1$ for each energy level when the Hamiltonian is in the $\mathcal{PT}$ exact phase. 
Below, we analytically show that the generalized parity indicator is well-defined in the $\mathcal{PT}$ exact phase, and the numerical verification of its property is provided in~\cite{SupMat}. 

In our ternary model, the $3\times 3$ Hamiltonian given by Eq.(\ref{MT_EffHam3}) respects $\mathcal{PT}$ symmetry and is a symmetric matrix, expressed as, 
\begin{equation}\label{AEQ_Transpose}
   \tilde{H}^T=\tilde{H}. 
\end{equation} 
Combining Eq.(\ref{AEQ_BiBasis}) and Eq.(\ref{AEQ_Transpose}), we obtain the relation that the $n$-th right and left eigenvectors satisfy:
\begin{equation}\label{SM_LeftRightVector}
\begin{split}
    \tilde{H}|\psi_n^R\rangle = E_n |\psi_n^R\rangle; \,\,\,\,
    \tilde{H}|\psi_n^L\rangle^{\ast} = E_n |\psi_n^L\rangle^{\ast}.
\end{split}
\end{equation}
It follows that:
\begin{equation}\label{SM_SymmetricEState}
   |\psi_n^R\rangle = e^{i\phi} |\psi_n^L\rangle^{\ast},
\end{equation}
where $\phi$ is an arbitrary phase factor and $|\psi_n^L\rangle^{\ast}$ is the complex conjugate of $|\psi_n^L\rangle$. 
In $\mathcal{PT}$ exact phase, the left and right eigenvectors are invariant under the $\mathcal{PT}$ operation, namely, 
\begin{equation}\label{SM_PTSymm}
    \mathcal{PT}|\psi_n^{R/L}\rangle
    =\mathcal{P}|\psi_n^{R/L}\rangle^{\ast}
    =e^{i\theta}|\psi_n^{R/L}\rangle,
\end{equation}
where the parity $\mathcal{P}$ is the unitary matrix, $\mathcal{P}\mathcal{P}^{\dagger}=1$, and satisfies
$\mathcal{P}^2=1$. 

Now we prove that the following quantity is real in $\mathcal{PT}$ exact phase, 
\begin{equation}\label{SM_Parity}
   p_n = \frac{\langle \psi_n^L|\mathcal{P}|\psi_n^R\rangle}{\langle \psi_n^L|\psi_n^R\rangle},
\end{equation}
where the denominator is the normalization factor of the biorthogonal eigenvector. 
Finally, it follows that 
\begin{equation}\label{SM_RealIndicator}
\begin{split}
p_n^{\dagger}&=\frac{\langle \psi_n^{R}|\mathcal{P}^{\dagger}|\psi_n^{L}\rangle}{\langle \psi_n^{R}|\psi_n^{L}\rangle} =\frac{\langle \psi_n^{L*}|\mathcal{P}^{\dagger}|\psi_n^{R*}\rangle}{\langle \psi_n^{L*}|\psi_n^{R*}\rangle}\\ & = \frac{\langle \psi_n^{L*}|\mathcal{P}^{\dagger}\mathcal{P}\mathcal{P}|\psi_n^{R*}\rangle}{\langle \psi_n^{L*}|\mathcal{P}^{\dagger}\mathcal{P}|\psi_n^{R*}\rangle}= \frac{\langle \psi_n^L|\mathcal{P}|\psi_n^R\rangle}{\langle \psi_n^L|\psi_n^R\rangle} = p_n, 
\end{split}
\end{equation}
where the second equality holds using Eq.(\ref{SM_LeftRightVector}), the third equality is valid using the involution relation $\mathcal{P}^2=1$, and the fourth equality holds using Eq.(\ref{SM_PTSymm}). 
Therefore, for our Hamiltonian, $p_n^{\dagger}=p_n$ is a real number in $\mathcal{PT}$ exact phase. 
The generalized parity indicator can be defined as the sign of $p_n$ in $\mathcal{PT}$ exact phase, that is, 
\begin{equation}\label{SM_GPI}
   \epsilon_n = \mathrm{sgn}(p_n),
\end{equation}
which is the same definition as Eq.(\ref{MT_ParityInd}) in the main text.

\section{Mathematical constraints of aEP and hEP for topological classification.}\label{AppendixC}
The mathematical constraints for aEP and hEP, which were used for defining complex vector field for topological characterizations, can be deduced by its distinct characteristics in the phase diagram.

To characterize the aEP, one can utilize its extremum characteristic in the phase diagram. 
\begin{equation}
    \partial_{\tilde{\gamma}}\tilde{\Delta}=0
\end{equation}
which leads to
\begin{equation}
    \partial_{\tilde{\gamma}}\tilde{\Delta}=-\partial_{\tilde{\gamma}}\mathrm{Disc}_E(f)/\partial_{\tilde{\Delta}}\mathrm{Disc}_E(f)=0
\end{equation}
and in turn requires 
\begin{equation}
\partial_{\tilde{\gamma}}\mathrm{Disc}_E(f)=0.
\end{equation}
Combining the condition of  $\mathrm{Disc}_E(f)=0$ and $\partial_{\tilde{\gamma}}\mathrm{Disc}_E(f)=0$, we obtain the function of aEP and hEP as: $\tilde{\Delta}=-1$ and $16\tilde{\Delta}^3+27\tilde{\Omega}^2(\tilde{\Delta}+1) =0$, respectively. 

%%%%%%%%%%%%%%%%%%%%%%%%  Supplementary Information  %%%%%%%%%%%%%%%%%%%%%%%%%%%%%%%
\let\oldaddcontentsline\addcontentsline
\renewcommand{\addcontentsline}[3]{}
\bibliography{refs_main}
\bibliographystyle{apsrev4-2}
\let\addcontentsline\oldaddcontentsline
\onecolumngrid
\newpage
\makeatletter
\renewcommand \thesection{S-\@arabic\c@section}
\renewcommand\thetable{S\@arabic\c@table}
\renewcommand \thefigure{S\@arabic\c@figure}
\renewcommand \theequation{S\@arabic\c@equation}
\makeatother
\setcounter{equation}{0}  
\setcounter{figure}{0}  
\setcounter{section}{0}  
{\begin{center}
	{\bf \large Supplemental Material for ``Topologically Protected Exceptional Points and Reentrant $\mathcal{PT}$ Phase in an Exact Ternary Model'' }
\end{center}}
\maketitle
% \tableofcontents
%%%%%%%%%%%%%%%%%%%%%%%%  Supplementary Information  %%%%%%%%%%%%%%%%%%%%%%%%%%%%%%%

\section{The phase boundary of the minimal model}
In this section, we will give the details of the method to obtain the phase boundary of the minimal model. 
First, we determine the phase boundary composed of the blue and green surfaces shown in Fig.~1(b) in the 3D phase diagram using the resultant method.
Then, in the 2D projected phase diagram shown in Fig.~1(c), we will prove that the phase boundaries of cyan and green lines are exactly two straight lines (namely independent of $\tilde{\Omega}$). 

\subsection{The phase boundaries in 3D phase diagram}
Here, we first give the solution of the blue surface in the 3D phase diagram in Fig.~1(b) using the discriminant method, where the real and imaginary parts of the Hamiltonian are the same. 
Then we further determine the green surface where the real parts of all three eigenvalues are equal, but the imaginary parts are not. 

\textbf{The blue surface}:
We begin with the characteristic equation of the Hamiltonian in Eq.(3) of the main text, which can be written as
\begin{equation}\label{SM_CharacEqua}
\begin{split}
     f(E, \tilde{\Delta}, \tilde{\Omega}, \tilde{\gamma}) 
     &= E^3 + c_2 E^2 + c_1 E + c_0 \\
     &=E^3-2 \tilde{\Delta} E^2+
    (\tilde{\Delta}^2+\tilde{\gamma}^2-\tilde{\Omega}^2/2-1)E +(\tilde{\Delta}+1)\tilde{\Omega}^2/2
    =0,
\end{split}
\end{equation}
where $c_{0,1,2}$ are the coefficients of the polynomial with respect to E and are functions of system parameters $\tilde{\Delta}, \tilde{\Omega}, \tilde{\gamma}$. 
For a given set of coefficients $c_2,c_1,c_0$, we can obtain corresponding solutions of $E$. 
Two of the three solutions will coincide when the coefficients satisfy the following equation, 
\begin{equation}\label{SM_DiscEqua}
     \mathrm{Disc}_{E}(f)= c_1^2c_2^2-4c_0c_2^3-4c_1^3 + 18 c_0 c_1 c_2 - 27 c_0^2=0,
\end{equation}
which is the discriminant~\cite{Discriminant_Notes} of the characteristic polynomial $f(E, \tilde{\Delta}, \tilde{\Omega}, \tilde{\gamma})$ regarding $E$. 
In the non-Hermitian case, Eq.(\ref{SM_DiscEqua}) gives us the 2D surface in 3D parameter space where at least two eigenvalues coincide. 
Therefore, the solutions of Eq.(\ref{SM_DiscEqua}) determine the blue surface shown in Fig.~1(b). 
When the Hamiltonian reduces to Hermitian limit ($\tilde{\gamma}=0$), the level-crossing points can be obtained, namely, the purple crease lying on the $\tilde{\gamma}=0$ plane shown in Fig.~1(b). 

\textbf{The green surface}:
For a given set of system parameters $\tilde{\Delta},\tilde{\Omega},\tilde{\gamma}$, the corresponding three eigenvalues can be obtained, which are generally complex. 
Here, we use $E_i$ and $E_i$ to represent the real and imaginary parts of the energy, respectively. 
According to the definition of the green surface, when the system is on the green surface, we can obtain three eigenvalues with the same real part but different imaginary parts. 
Since the Hamiltonian has $\mathcal{PT}$ symmetry, the system eigenvalues have always one real energy and two complex energies that are complex conjugate with each other. 
Under these considerations, we obtain the condition for the green surface as follows,
\begin{equation}\label{SM_GreenSurf}
\begin{split}
     &f(E_r, \tilde{\Delta}, \tilde{\Omega}, \tilde{\gamma}) =f(E_r+i E_i, \tilde{\Delta}, \tilde{\Omega}, \tilde{\gamma})=f(E_r-i E_i, \tilde{\Delta}, \tilde{\Omega}, \tilde{\gamma})=0; \,\,\,\, E_i \neq 0,
\end{split}
\end{equation}
where $E_r,E_i,\tilde{\Delta}, \tilde{\Omega}, \tilde{\gamma}$ are real variables. 
The solution satisfying Eq.(\ref{SM_GreenSurf}) in ($\tilde{\Delta}, \tilde{\Omega}, \tilde{\gamma}$) parameter space is exactly the green surface in Fig.~1(b). 

Next, we will solve the first three equations in Eq.(\ref{SM_GreenSurf}) using the resultant method~\cite{woody2016polynomial}. 
Note that the last two equations, $f(E_r+i E_i, \tilde{\Delta}, \tilde{\Omega}, \tilde{\gamma})=0$ and $f(E_r-i E_i, \tilde{\Delta}, \tilde{\Omega}, \tilde{\gamma})=0$, are not independent because the Hamiltonian respects the $\mathcal{PT}$ symmetry.
Therefore, the green surface is determined by the following three independent real equations
\begin{equation}\label{SM_GreenSurfv1}
\begin{split}
     &f(E_r, \tilde{\Delta}, \tilde{\Omega}, \tilde{\gamma}) =0; \\
     &\mathrm{Re}f(E_r+i E_i, \tilde{\Delta}, \tilde{\Omega}, \tilde{\gamma}) =0; \\
     &\mathrm{Im}f(E_r+i E_i, \tilde{\Delta}, \tilde{\Omega}, \tilde{\gamma}) =0,
\end{split}
\end{equation}
and the nonzero imaginary part
\begin{equation}\label{SM_GreenSurfv2}
    E_i\neq 0.
\end{equation}
The set of equations in Eq.(\ref{SM_GreenSurfv1}) can be solved by the resultant method. 
We can obtain two equations by eliminating the variable $E_r$ using the resultant method, namely, 
\begin{equation}\label{SM_GreenSurfv3}
\begin{split}
     \mathrm{Res}[f(E_r, \tilde{\Delta}, \tilde{\Omega}, \tilde{\gamma}), \mathrm{Re}f(E_r+i E_i, \tilde{\Delta}, \tilde{\Omega}, \tilde{\gamma}), E_r]
     =E_i^6 \, g_1(E_i, \tilde{\Delta}, \tilde{\Omega}, \tilde{\gamma})=0; \\
     \mathrm{Res}[f(E_r, \tilde{\Delta}, \tilde{\Omega}, \tilde{\gamma}), \mathrm{Im}f(E_r+i E_i, \tilde{\Delta}, \tilde{\Omega}, \tilde{\gamma}), E_r]
     =E_i^3 \, g_2(E_i, \tilde{\Delta}, \tilde{\Omega}, \tilde{\gamma})=0,
\end{split}
\end{equation}
Then we can get rid of the variable $E_i$ and obtain the solution in ($\tilde{\Delta}, \tilde{\Omega}, \tilde{\gamma}$) parameter space 
\begin{equation}\label{SM_GreenSurfv4}
\begin{split}
     g(\tilde{\Delta}, \tilde{\Omega}, \tilde{\gamma}) &:=\mathrm{Res}[g_1(E_i, \tilde{\Delta}, \tilde{\Omega}, \tilde{\gamma}), g_2(E_i, \tilde{\Delta}, \tilde{\Omega}, \tilde{\gamma}), E_i] \\
     &=(-36 \tilde{\Delta} + 36 \tilde{\gamma}^2 \tilde{\Delta} + 4 \tilde{\Delta}^3 + 27 \tilde{\Omega}^2 + 9 \tilde{\Delta} \tilde{\Omega}^2)^6/64=0,
\end{split}
\end{equation}
which is a 2D surface in the 3D parameter space. 
% Note that here $g(\tilde{\Delta}, \tilde{\Omega}, \tilde{\gamma})=0$ includes the solutions of $E_i=0$.  
% Note that when $E_i=0$, the three equations in Eq.(\ref{SM_GreenSurfv1}) are no longer independent of each other, and the resultant method is invalid. 
Combining with the constraint Eq.(\ref{SM_GreenSurfv2}), $E_i\neq 0$, we can finally find determine the green surface, which must be in the $\mathcal{PT}$ broken phase, namely, above the blue surface as shown in Fig.~1(b). 

\subsection{The phase boundaries in 2D projected phase diagram}

Here, we give the 2D projection phase diagram, as shown in Fig.~1(c). 
The projection of the 3D phase diagram on the $\tilde{\Delta}-\tilde{\Omega}$ plane is divided into seven regions by four special phase boundaries, marked as purple, red, cyan, and green curves in Fig.~1(c). 

The purple curve is the intersection of the blue surface and the $\tilde{\gamma}=0$ plane in Fig.~1(b). 
Therefore, when $\tilde{\gamma}=0$, Eq.(\ref{SM_DiscEqua}) yields the purple curve, that is, 
\begin{equation}\label{SM_PurpleCurve}
    \tilde{\Omega}^2 - 4\tilde{\Delta} - 4 = 0.
\end{equation}
When the system crosses this phase boundary in the $\tilde{\gamma}=0$ plane, the two energy levels cross and reopen, accompanied by a flip of the corresponding parity indicators. 
One example is shown in Fig.~2(b) and (c). 
The Hamiltonian with $\tilde{\gamma}=0$ in Fig.~2(b) lies in the $B1$ region of the 2D projected parameter space, and its three energy levels have parity indicators ``+ - +" in order of energy from highest to lowest. 
As the system crosses the purple curve and goes into the $B2$ region, the three energy levels have parity ``- + +" in order of energy from highest to lowest, marked in Fig.~2(c). 
It shows that the purple curve is a phase boundary. 

The red curve in Fig.~1(c) is the projection of the red crease of the blue surface (called $\mathcal{PT}$ phase boundary) in Fig.~1(b). 
The red crease is the intersection between the blue and the green surfaces in Fig.~1(b). 
Therefore, the red curve in Fig.~1(c) can be obtained using the resultant method. 
Combining Eq.(\ref{SM_DiscEqua}) and Eq.(\ref{SM_GreenSurfv4}), we can eliminate $\tilde{\gamma}$ using the resultant method and obtain the equation of the red curve in Fig.~1(c), 
\begin{equation}\label{SM_RedCurve}
    16 \tilde{\Delta}^3 + 27 \tilde{\Omega}^2 
    + 27 \tilde{\Delta} \tilde{\Omega}^2 = 0.
\end{equation}
When the system is in the reentry region, with $\tilde{\gamma}$ increasing, the system will cross the $\mathcal{PT}$ phase boundary three times.  
Actually, the red curve is one boundary of the reentry phase region. 
When the system is on the red curve, as $\tilde{\gamma}$ increases, the system will experience a third-order EP, where the three eigenvalues of the Hamiltonian are the same, and the three eigenstates are parallel to each other.
One example is shown in Fig.~2(d). 
It shows that the red curve is a phase boundary. 

The cyan line in Fig.~1(c) is another boundary of the reentry region, where the system will encounter the a-EP as $\tilde{\gamma}$ increases. 
Here, we prove that the cyan line is exactly a straight line, which means that the phase boundary is independent of $\tilde{\Omega}$.
From a slice of the 3D phase diagram shown in Fig.~2(a), one can see that the slope of the PT phase boundary (the blue surface in Fig.~1(b)) at the a-EP point is zero. 
Therefore, the phase boundary satisfies both of the following equations,
\begin{equation}\label{SM_CyanCondition}
   \mathrm{Disc}_{E}(f)=0; \,\,\,\, 
   \partial_{\tilde{\gamma}}\mathrm{Disc}_{E}(f)=0,
\end{equation}
where $b(\tilde{\Delta}, \tilde{\Omega}, \tilde{\gamma})=0$ is the equation of the blue surface defined in Eq.(\ref{SM_DiscEqua}). 
Using the resultant method, we obtain the solution to the two equations. 
The solution consists of three curves, two of which happen to be Eq.(\ref{SM_PurpleCurve}) and Eq.(\ref{SM_RedCurve}) because they correspond to the two creases of the blue surface, and the remaining one is 
\begin{equation}\label{SM_CyanCurve}
   \tilde{\Delta}=-1,
\end{equation}
which is a straight line as the left boundary of the reentry phase region, as shown in Fig.~1(c). 

Now we examine the boundary of the region $C$ and $D$ in Fig.~\ref{fig:s1}(c). 
When the system lies in the region $C$, with $\tilde{\gamma}$ increasing, the system first crosses the $\mathcal{PT}$ phase boundary and enters the $\mathcal{PT}$ broken phase, then three real parts coincide on the green surface, as shown in Fig.~1(b). 
While in the region $D$, the system experiences once $\mathcal{PT}$ phase boundary as $\tilde{\gamma}$ increases. 
Therefore, the boundary between region $C$ and $D$ is the limitation of the green surface on the $\tilde{\Delta}-\tilde{\Omega}$ plane as $\tilde{\gamma}$ tends to infinity. 
From the equation of the green surface in Eq.(\ref{SM_GreenSurfv4}), we have 
\begin{equation}\label{SM_GreenBound}
   \tilde{\gamma}^2 = \frac{36 \tilde{\Delta} - 4\tilde{\Delta}^3-27\tilde{\Omega}^2-9\tilde{\Delta}\tilde{\Omega}^2 }{36 \tilde{\Delta}}.
\end{equation}
Clearly, for any fixed $\tilde{\Omega}$ (or any given $\tilde{\Omega}$ slice of the 3D phase diagram), as $\tilde{\gamma}$ goes to infinity, the solution of the boundary is 
\begin{equation}\label{SM_GreenLine}
   \tilde{\Delta} = 0,
\end{equation}
which is the green straight line in Fig.~1(c) and independent of $\tilde{\Omega}$. 

\section{The excitonic fraction in different sections for Fig.2 of the main text}

\begin{figure}[t]
\begin{center}
\includegraphics[width=1\linewidth]{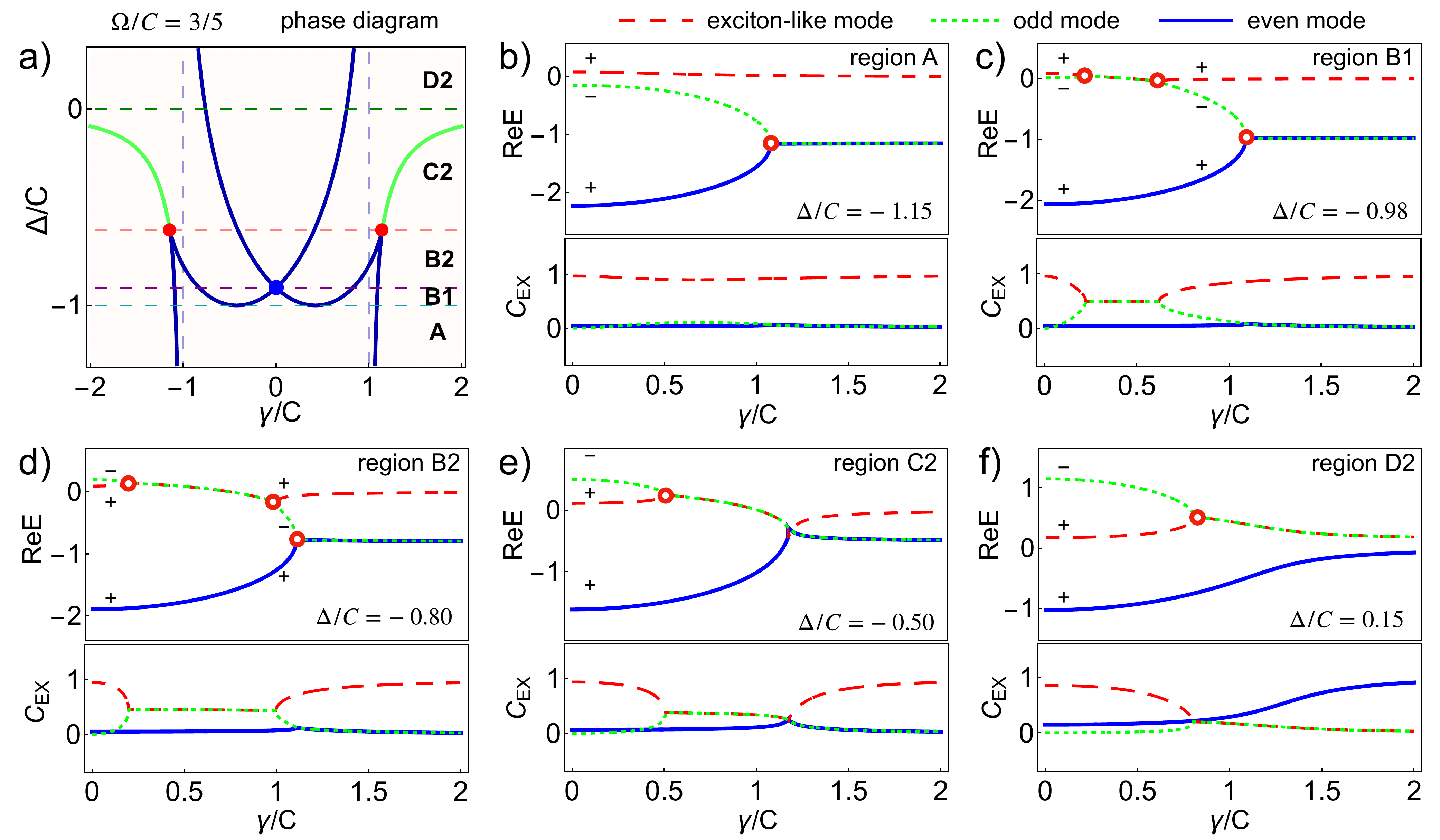}
\par\end{center}
\protect\caption{\label{fig:s1} 
    The excitonic fraction in different sections for Fig.~2 of the main text.} 
\end{figure}

Here, we present the excitonic fraction in different phase sections as a supplement of Fig.~2 in the main text. 
Our $\mathcal{PT}$-symmetric ternary Hamiltonian in the basis $\{|A\rangle,|B\rangle,|C\rangle\}$ reads 
\begin{equation}
    \begin{pmatrix}
    \tilde{H}=\tilde{\Delta}-i \tilde{\gamma} & -1 &\tilde{\Omega}/2 \\
    -1 & \tilde{\Delta}+i \tilde{\gamma} & \tilde{\Omega}/2\\
    \tilde{\Omega}/2 & \tilde{\Omega}/2 & 0
    \end{pmatrix}.
\end{equation}
The basis $|A\rangle,|B\rangle,|C\rangle$ represent the even, odd cavity modes, and exciton-like mode, respectively.
Based on this, the excitonic fraction for a given eigenstate $\psi_{n}$ with energy $E_n$ is given by:
\begin{equation}
    C_{\mathrm{EX}} = |\langle C|\psi_n\rangle|^2.
\end{equation}
We evaluate the excitonic fraction for all eigenstates in different five phase regions shown in Fig.~\ref{fig:s1}(a) and plot them in Fig.~\ref{fig:s1}(b)-(f). 
Note that in region $B1$ and $B2$ exciton is strongly coupled with the cavities throughout not only the two $\mathcal{PT}$ exact phases, but also the first $\mathcal{PT}$ broken phase, where the two polariton modes are degenerate.

\section{Numerical verification of generalized parity indicator}

\begin{figure}[t]
\begin{center}
\includegraphics[width=1\linewidth]{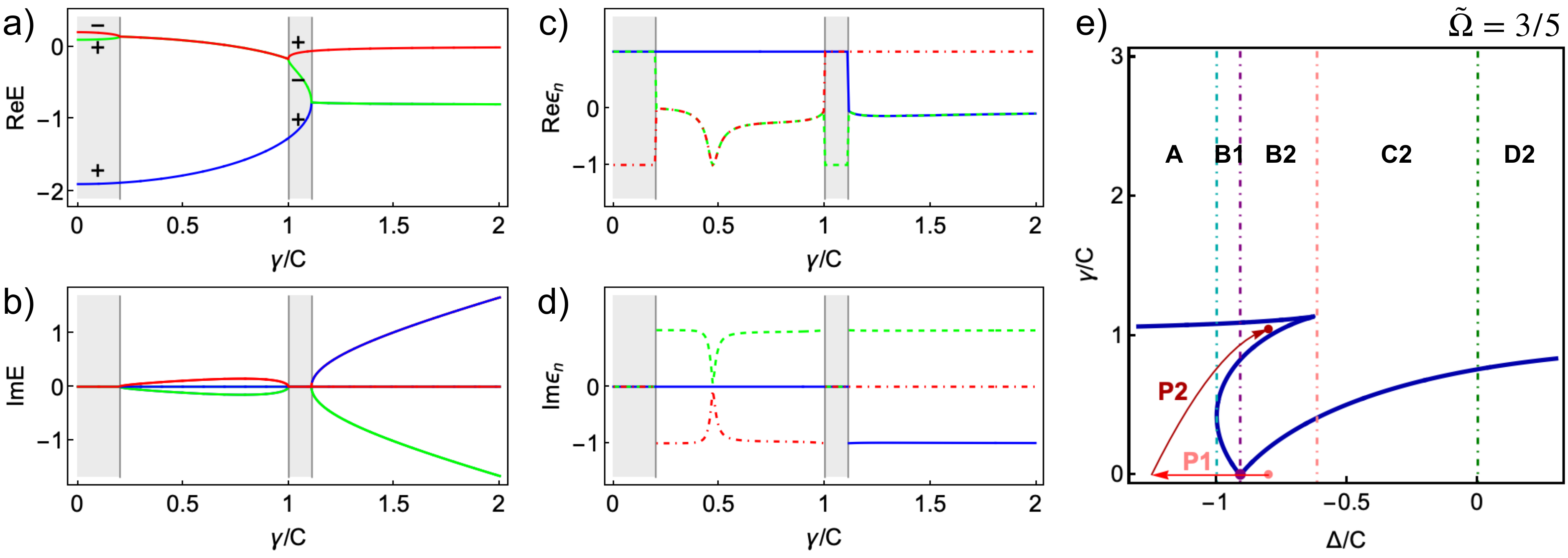}
\par\end{center}
\protect\caption{\label{fig:s2} 
    The numerical verification for the generalized parity indicator. The system parameters are set to be $(\tilde{\Delta},\tilde{\Omega})=(-4/5,3/5)$. The evolutions of eigenvalues with $\tilde{\gamma}$ are shown in (a)(b), and corresponding generalized parity indicators are calculated in (c)(d). The eigenvalues and their generalized parity indicator $\epsilon_n$ are marked in the same color. The evolution paths $\mathrm{P}1$ and $\mathrm{P}2$ in $\tilde{\Delta}-\tilde{\Omega}$ plane are indicated in (e). } 
\end{figure}

In this section, we numerically verify that the generalized parity indicator indeed works. 
The system parameters are set to be $(\tilde{\Delta},\tilde{\Omega})=(-4/5,3/5)$. 
The eigenvalues and eigenvectors of the Hamiltonian evolve with $\tilde{\gamma}$. 
We show in Fig.~\ref{fig:s1}(a)(b) the evolution of the real and imaginary parts of the eigenvalues with $\tilde{\gamma}$. 
In the beginning, the system is in the Hermitian limit and then enters the first $\mathcal{PT}$ entry region as $\tilde{\gamma}$ increases. 
In the weak non-Hermiticity region ($\tilde{\gamma} \sim 0$), the generalized parity indicator can be calculated and shown in Fig.~\ref{fig:s1}(c)(d). 
The three eigenvalues and their generalized parity indicator $\epsilon_n$ are marked in the same color. 
Therefore, in order of energy (from highest to lowest), the corresponding generalized parity indicators are ``- + +". 
Note that 
(i) these generalized parity indicators reduce to the parity indicators (the expectation value of $\mathcal{P}$) in the Hermitian limit; 
(ii) Two modes with opposite generalized parity indicators merge and form EP. 

It's worth noting that in the strong non-Hermiticity region ($\tilde{\gamma}\gg 0$), the system reenters the $\mathcal{PT}$ exact phase as shown in Fig.~\ref{fig:s1}(a)(b), and the generalized parity indicator still works. 
In this region, the generalized parity indicators in order of energy are calculated as ``+ - +". 
Note that the first two parity indicators are flipped in these two $\mathcal{PT}$ exact phases. 
It can be understood from the phase diagram in Fig.~\ref{fig:s1}(e). 
The system is initially at the pink dot in the $\tilde{\Delta}-\tilde{\Omega}$ plane of Fig.~\ref{fig:s1}(e), and finally evolves into the dark red dot. 
The evolution path is composed of $\mathrm{P}1$ and $\mathrm{P}2$ arrows as illustrated in Fig.~\ref{fig:s1}(e). 
Since the $\mathrm{P}1$ path (in the Hermitian limit) crosses the level-crossing point (the purple dot), the parity indicators of two energy levels are flipped. 
While along the $\mathrm{P}2$ evolution path, the generalized parity indicators remain unchanged. 
The system can adiabatically evolve along the $\mathrm{P}2$ path because the system along this path retains non-degenerate real eigenvalues and does not cross any phase boundary. 
That is the reason why two generalized parity indicators are flipped in the first and second $\mathcal{PT}$ exact phases, as illustrated in Fig.~\ref{fig:s1}(a)-(d). 
\begin{figure}[t]
\begin{center}
\includegraphics[width=1\linewidth]{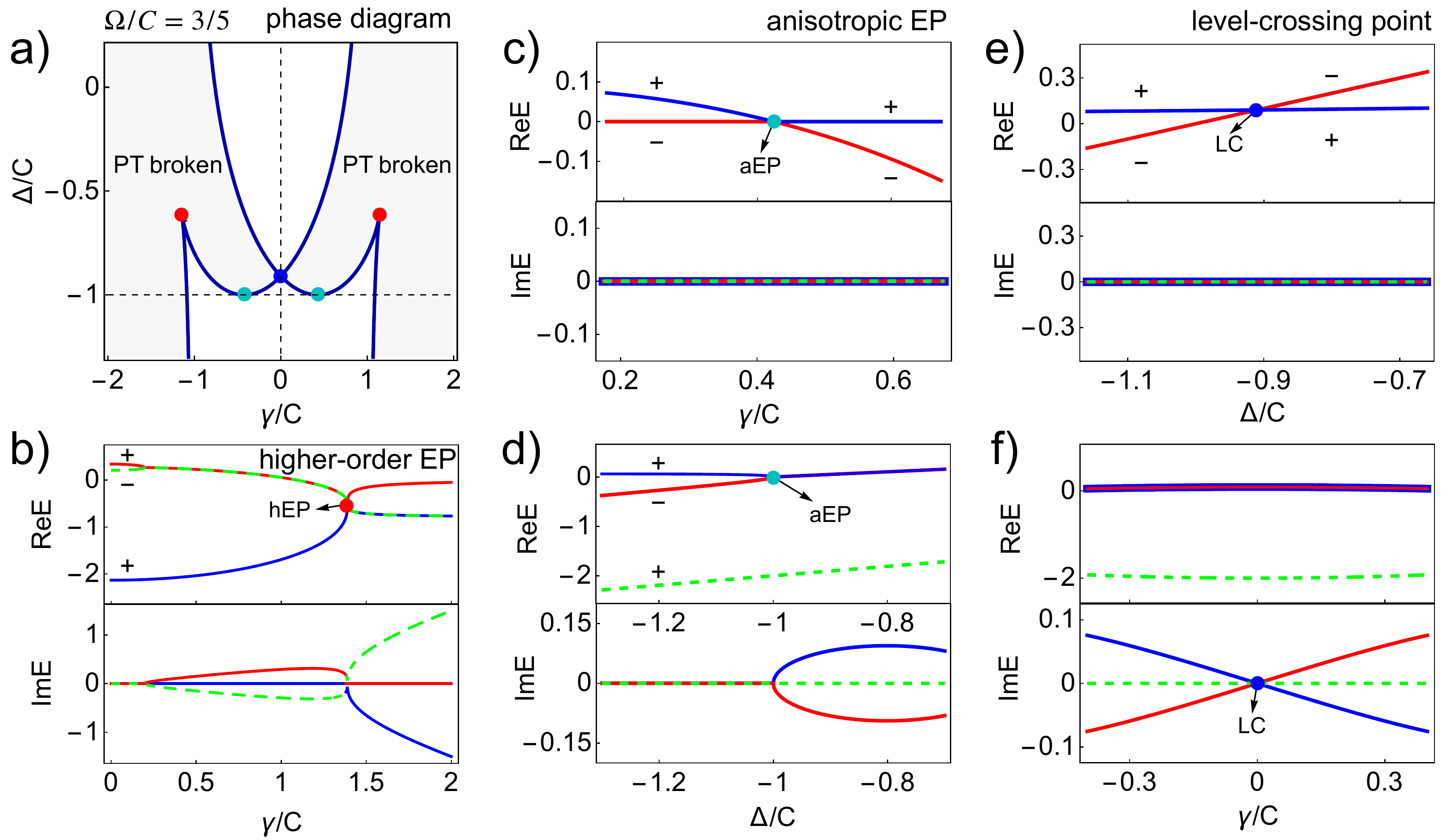}
\par\end{center}
\protect\caption{\label{fig:s3} 
    Higher-order EP and near-zero-gain. The system parameters are set to be $\{\tilde{\Delta},\tilde{\Omega}\}=\{-3/4,1\}$. The evolutions of real and imaginary parts of eigenvalues with $\tilde{\gamma}$ are shown in (a) and (b), respectively. The sensitivity of the real energy to $\tilde{\gamma}$ near EP3 is shown in (c), where the analytical result (pink curve) agrees with the numerical result (blue dashed curve), i.e., $\delta \mathrm{Re}E \sim \delta\tilde{\gamma}^{1/3}$. } 
\end{figure}

% \section{Higher-order EP and near-zero-gain $\mathcal{PT}$ symmetry breaking in ternary polaritonic non-Hermitian system}
\section{Higher-order EPs and anisotropic EPs in ternary $\mathcal{PT}$ symmetric non-Hermitian system}

In this section we discuss the two important consequences of polaritonic ternary non-Hermitian model, higher-order EP and near-zero-gain $\mathcal{PT}$ at LC line. 

The LC line separates regions 1 and 2, characterized by different mode parities,(Fig.~\ref{fig:s1}) such as $B1$ and $B2$ (Fig.~\ref{fig:s1}(c)-(d)). 
We mark it by the blue dot in Fig.~\ref{fig:s1}(a) and Fig.~\ref{fig:s3}(a). 
It is the intersection of the lower EP surface I and the middle EP surface II (Fig. 1(a)). It lies on the $\tilde{\gamma}=0$ plane, corresponding to the Hermitian limit where Eq.(\ref{SM_DiscEqua}) gives the condition of level crossing (LC): $\tilde{\Omega}^2-4\tilde{\Delta}-4=0$. The energy level crossing of two highest levels with opposite parities can be seen by evolving the system along the path on the $\tilde{\gamma}=0$ plane with crossing LC point.(Fig.~\ref{fig:s3}(e))
The LC point gives rise to another interesting feature, a near-zero-gain $\mathcal{PT}$ broken phase. While the LC corresponds to degeneracy points instead of EP, any finite $\tilde{\gamma}$ near LC drives the system immediately to the $\mathcal{PT}$ broken phase, which may be utilized for realizing ultra-low threshold photonic devices.
\section{Sensitivity analysis of higher-order EP in the non-Hermitian polariton system}

\begin{figure}[t]
\begin{center}
\includegraphics[width=0.9\linewidth]{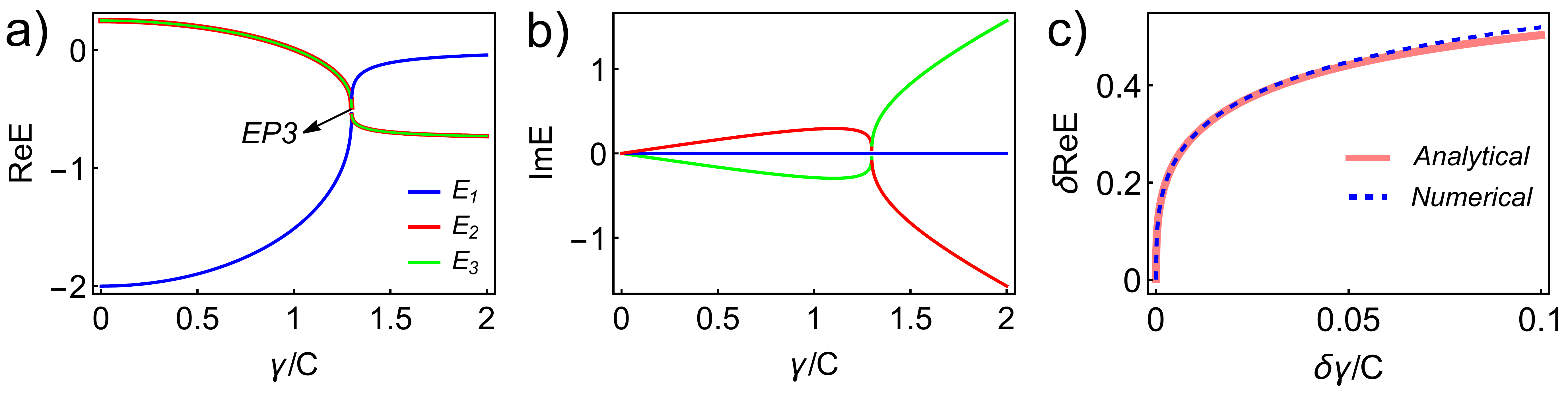}
\par\end{center}
\protect\caption{\label{fig:s4} 
    The numerical verification of sensitivity of EP3. The system parameters are set to be $\{\tilde{\Delta},\tilde{\Omega}\}=\{-3/4,1\}$. The evolutions of real and imaginary parts of eigenvalues with $\tilde{\gamma}$ are shown in (a) and (b), respectively. The sensitivity of the real energy to $\tilde{\gamma}$ near EP3 is shown in (c), where the analytical result (pink curve) agrees with the numerical result (blue dashed curve), i.e., $\delta \mathrm{Re}E \sim \delta\tilde{\gamma}^{1/3}$. } 
\end{figure}

This section will numerically verify the $\delta E \propto \delta \tilde{\gamma}^{1/3}$ sensitivity around the EP3. 
As an example, the system parameters are chosen to be $\{\tilde{\Delta},\tilde{\Omega}\}=\{-3/4,1\}$, and the Hamiltonian with $\tilde{\gamma}$ becomes 
\begin{equation}\label{SM_EP3ham}
   \tilde{H} = 
   \begin{pmatrix}
   -i \tilde{\gamma} - 3/4 & -1 & 1/2 \\    
   -1 & i \tilde{\gamma} - 3/4 & 1/2 \\
   1/2 & 1/2 & 0
   \end{pmatrix}.
\end{equation}
When $\tilde{\gamma}=\tilde{\gamma}_{\mathrm{EP3}}=3\sqrt{3}/4$, the system is at EP3 where three eigenvalues of the Hamiltonian are $E_{\mathrm{EP3}}=1/2$, and the corresponding three eigenstates are parallel to each other. 
We will examine the sensitivity of energy to the change of $\delta \tilde{\gamma}$ around $\tilde{\gamma}_{\mathrm{EP3}}$. 
The characteristic equation dependent of $\delta \tilde{\gamma}$ reads
\begin{equation}\label{SM_CharacEquaEP3}
   E^3 + \frac{3E^2}{2} + (\delta \tilde{\gamma}^2+\frac{3\sqrt{3}}{2}\delta \tilde{\gamma}+\frac{3}{4})E + \frac{1}{8}=0. 
\end{equation}
This equation can be perturbatively expanded using a Newton–Puiseux series.
The energy can be approximated into, $E\sim E_{\mathrm{EP3}} + c_1 \delta \tilde{\gamma}^{1/3} + c_2 \delta \tilde{\gamma}^{2/3}$, with the coefficients $c_1$ and $c_2$ being complex in general. 
After substituting the approximated $E$, the characteristic equation becomes
\begin{equation}\label{SM_CharacEquaEP3v1}
   \left(\frac{3 \sqrt{3}}{4}-c_1^3\right) x+\left(-3 c_1^2 c_2-\frac{3 \sqrt{3} c_1}{2}\right) x^{4/3} +
   \left(-3 c_1 c_2^2-\frac{3 \sqrt{3} c_2}{2}\right) x^{5/3} + \cdots=0. 
\end{equation}
Let the coefficients of the first two terms be zero, and we obtain three sets of values for the coefficients $c_1$ and $c_2$. 
Correspondingly, the three eigenvalues can be expanded as
\begin{equation}\label{SM_EP3Eigens}
\begin{split}
    &E_1 = E_{\mathrm{EP3}} + \frac{\sqrt{3}}{2^{2/3}} \delta \tilde{\gamma}^{1/3} - \frac{1}{2^{1/3}} \delta \tilde{\gamma}^{2/3}; \\ 
    &E_2 = E_{\mathrm{EP3}} - \frac{(-1)^{1/3} \sqrt{3}}{2^{2/3}} \delta \tilde{\gamma}^{1/3} - \frac{(-1)^{2/3}}{2^{1/3}} \delta \tilde{\gamma}^{2/3}; \\ 
    &E_3 = E_{\mathrm{EP3}} + \frac{(-1)^{2/3} \sqrt{3}}{2^{2/3}} \delta \tilde{\gamma}^{1/3} + \frac{(-1)^{1/3}}{2^{1/3}} \delta \tilde{\gamma}^{2/3}.
\end{split}
\end{equation}
Note that the EP3 splits into $\mathcal{PT}$ broken phase after adding the perturbation $\delta \tilde{\gamma}$, thus $\mathrm{Re}E_2=\mathrm{Re}E_3$, as shown in Fig.~\ref{fig:s4}(a)(b). 

Based on the above analysis, we calculate the sensitivity order as 
\begin{equation}\label{SM_EP3Sensitivity}
    \delta \mathrm{Re}E = \mathrm{Re}[E_1-E_2] \sim \frac{3\sqrt{3}}{4} \, 2^{1/3}\, \delta\tilde{\gamma}^{1/3},
\end{equation}
which agrees well with the numerical results shown in Fig.~\ref{fig:s4}(c). 

% \section{Mathematical constraints for Anisotropic and Higher-order Exceptional points}

% The mathematical constraints for aEP and hEP, which were used for defining complex vector field for topological characterizations, can be deduced by its distinct characteristics in the phase diagram.

% To characterize the aEP, one can utilize its extremum characteristic in the phase diagram. (Blue dots in Fig. S3(a)) 
% \begin{equation}
%     \partial_{\tilde{\gamma}}\tilde{\Delta}=0
% \end{equation}
% which leads to
% \begin{equation}
%     \partial_{\tilde{\gamma}}\tilde{\Delta}=-\partial_{\tilde{\gamma}}\mathrm{Disc}_E(f)/\partial_{\tilde{\Delta}}\mathrm{Disc}_E(f)=0
% \end{equation}
% and in turn requires 
% \begin{equation}
% \partial_{\tilde{\gamma}}\mathrm{Disc}_E(f)=0.
% \end{equation}
% Combining the condition of  $\mathrm{Disc}_E(f)=0$ and $\partial_{\tilde{\gamma}}\mathrm{Disc}_E(f)=0$, we obtain the function of aEP and hEP as: $\tilde{\Delta}=-1$ and $16\tilde{\Delta}^3+27\tilde{\Omega}^2(\tilde{\Delta}+1) =0$, respectively. 


\end{document}